\documentclass[editedvolume,numreferences]{crckbked}

\usepackage{epsfig}
\usepackage{amsmath,amssymb}

\begin{document}

\bibliographystyle{klunum}

\tableofcontents

\begin{article}
\def\alt{\lsim}
\def\agt{\gsim}

\begin{opening}
\title{Photon Assisted Tunneling in Quantum Dots}
\runningtitle{Photon Assisted Tunneling in Quantum Dots}

\author{W.G. \surname{van der Wiel}, T.H. \surname{Oosterkamp},\\
S. \surname{De Franceschi}, C.J.P.M. \surname{Harmans}\\ and L.P.
\surname{Kouwenhoven}} \runningauthor{W.G. \surname{van der Wiel}
{\it et al.}}

\institute{Department of Applied Physics and DIMES, Delft
University of Technology, PO Box 5046, 2600 GA Delft, The
Netherlands}

\end{opening}

\noindent We review experiments on single-electron transport
through single quantum dots in the presence of a microwave signal.
In the case of a small dot with well-resolved discrete energy
states, the applied high-frequency signal allows for inelastic
tunnel events that involve the exchange of photons with the
microwave field. These photon assisted tunneling (PAT) processes
give rise to sideband resonances in addition to the main
resonance. Photon absorption can also lead to tunneling via
excited states instead of tunneling via the ground state of the
quantum dot. The manipulation of quantum dots by a microwave field
is an important ingredient for the possible application of quantum
dots as solid state quantum bits, which forms a motivation for
this review.

\section{Introduction}

Transport properties of quantum dots at zero frequency have been
extensively studied and by now many aspects are well understood
\cite{Coulomb}. In some studies finite, but still low frequency
signals were applied to a gate electrode nearby the quantum dot.
Capacitance spectroscopy on quantum dots has been performed at kHz
frequencies \cite{Ashoori}. At MHz frequencies, quantum dots can
be operated as turnstiles or pumps \cite{Leopump}. These
frequencies, $f$, are low in the sense that the photon energy,
$hf$, is much smaller than the thermal energy, $k_{B}T,$ and thus
the discrete photon character cannot be discerned. For
sufficiently high frequencies, such that $ hf\gg k_{B}T$, the
interaction of the electromagnetic field with the electrons
confined in a quantum dot would be analogous to light spectroscopy
studies on atoms. However, different quantum dots are not
microscopically equal. The response of an ensemble of quantum dots
to light excitation is therefore strongly averaged. Despite this
averaging, excitation studies on arrays of quantum dots by
far-infrared light (i$.$e$.$ the THz regime) have revealed the
spectrum of collective modes \cite{Merkt}, i$.$e$.$ the sloshing
modes of the whole electron puddle in the external potential.
Excitations within the electron puddle are difficult to create
since, according to the generalized Kohn theorem \cite{Merkt}, the
dipole field of far-infrared radiation does not couple to the
relative coordinates of electrons confined in a parabolic quantum
dot. This problem is circumvented by inelastic light scattering
experiments which have been able to detect electronic excitations
in arrays of quantum dots that can be related to a discrete
single-particle spectrum \cite{Raman}. Recently, this technique
has also probed excitons in a single quantum dot \cite{Gerhard}.

In this review, we discuss electron transport experiments on
single quantum dots, that are irradiated by a microwave signal. In
contrast to light transmission, luminescence, or inelastic light
scattering measurements (see for a review Ref$.$
\cite{Hawrylak-book}), we measure the dc current in response to a
microwave signal. Current can flow through a quantum dot when a
discrete energy state is aligned to the Fermi energies of the
leads. This resonant current is carried by elastic tunneling of
electrons between the leads and the dot. An additional
time-varying potential $\widetilde{V}\cos (2\pi ft)$ can induce
{\it inelastic} tunnel events when electrons exchange photons of
energy $hf$ with the oscillating field. This inelastic tunneling
with discrete energy exchange is known as photon assisted
tunneling (PAT). Microwave studies have a long tradition in the
field of superconductor-insulator-superconductor tunnel junctions
\cite{Tucker}, for which the theory was first described by Tien
and Gordon in 1963 \cite{Tien}. Despite many proposals and a long
search \cite{failures}, it took thirty years before PAT was also
observed in a non-superconducting system. In 1993 PAT features
were seen in the current-voltage characteristics of a GaAs/AlGaAs
superlattice under THz irradiation from a free-electron laser
\cite{SantaBarbara}. Starting in 1994, PAT was also found in
experiments on single-electron transport through semiconductor
quantum dots [12-16]. 
The quantum dots in Ref$.$ \cite{Leo-PAT2} were rather large
and effectively had a continuous density of states. Here, we focus
on PAT processes through quantum dots with well-resolved, discrete
energy states. For these small dots, the resonant tunneling peak
in the current develops photon sideband resonances when we apply
microwaves \cite{Tjerk PAT}. The energy separation between main
peak and sidebands can be used as a spectroscopic measurement of
the energy levels in the dot.

The relevant ac regime for quantum dots is at much lower frequency
than visible light. We list the important frequency scales in
Table I. The single-particle level spacing $\Delta \varepsilon $
is 0.05 - 0.5 meV for typical dots and the charging energy,
$e^{2}/C$, is usually 0.2 - 2 meV. To observe effects from a
finite $\Delta \varepsilon $ and $e^{2}/C$, these energies should
exceed the thermal broadening $\sim 4k_{B}T$. Other characteristic
frequencies of the dot are related to the transport times. $\Gamma
$ is the typical rate to tunnel on or off the dot, which can be
arbitrarily small for opaque tunnel barriers. This frequency is
set by the transmission coefficient of the barriers and should be
kept smaller than $\Delta \varepsilon$ otherwise the level
broadening exceeds the spacing between the single-particle states.
The final time scale is the tunneling time; i$.$e$.$ the actual
time spent during tunneling through the barrier. This time is
quite short ($\sim $2 ps) for typical barriers (calculated within
the B\"{u}ttiker-Landauer framework \cite{tunnel-time}). To access
these time scales, ac signals can be applied, and the effects on
the dc transport can be measured.

\begin{table}[htbp]
      \begin{tabular}[t]{|c|c|c|}
\hline
{\bf Quantity} & {\bf Equivalent frequency} & {\bf Typical frequencies} \\
\hline Thermal broadening & $\sim 4k_{B}T/h$ & 10 GHz (at 100 mK)
\\ \hline $
\begin{array}{c}
\text{Tunneling rate} \\
\text{on/off the dot}
\end{array}
$ & $\Gamma $ & 0 - 100 GHz \\ \hline $
\begin{array}{c}
\text{Level spacing (or} \\
\text{inverse traversal time)}
\end{array}
$ & $\Delta \varepsilon /h$ & 10 - 100 GHz \\ \hline Charging
energy & $e^{2}/hC$ & 40 - 400 GHz \\ \hline Tunneling time &
$1/\tau _{tunnel}$ & 200 GHz - 1 THz \\ \hline
\end{tabular}
    \caption{A list of the important energy/frequency
    scales for transport through quantum dots. For $f$ = 10 GHz the
    photon energy, $hf$, is 40 $\mu $eV.}
      \label{tab:1}
  \end{table}

If $f$ $\ll \Gamma $ each electron sees an essentially static
potential and we are in the adiabatic regime
\cite{Ashoori,MarcusPump}. If $f$ $\gg \Gamma $, each electron
experiences many cycles of the ac signal while it is on the dot;
i$.$e$.$ the non-adiabatic regime. If $hf\ll 4k_{B}T$, single
photon processes are masked by thermal fluctuations, and a
classical description is appropriate \cite{Leopump}. Thus, the
discreteness of the photon energy can be observed in the
non-adiabatic, high-frequency regime: $hf\gg h\Gamma ,4k_{B}T$.
This is the quantum, or time-dependent regime for which it is
essential to solve the time-dependent Schr\"{o}dinger equation for
the tunneling electron.

This review is divided into two parts. First, in section
\ref{theory}, we discuss the theory of PAT through a single
junction (\ref{PATsing}) and calculations of PAT through a single
quantum dot using a master equation approach (\ref{master}).
Numerical PAT calculations are presented in sections
\ref{numresults} and \ref{PAP}. Second, in section
\ref{experiments}, we discuss experimental results of PAT
measurements on single lateral quantum dots. The sample geometry
is discussed in section \ref{sample}, the sensitivity of pumping
to the applied microwave frequency in \ref{frequency}. PAT in the
low ($hf<\Delta \varepsilon $) and high frequency ($hf>\Delta
\varepsilon $) regime are discussed in \ref{PATlowfreq} and
\ref{PAThighfreq}, respectively. In sections \ref{freqdep} through
\ref{Pdep} results are given of the dependence of PAT on microwave
frequency, magnetic field and microwave power, respectively. We
conclude and discuss the results in section \ref{conclusions}.

\section{Theory}
\label{theory}

\subsection{PAT through a single junction}
\label{PATsing}

First, we briefly outline photon assisted transport through a
single tunnel junction separating two metallic leads. An
oscillating potential difference across a junction,
$\widetilde{V}\cos (2\pi ft)$, where $\widetilde{V}$ is the ac
amplitude, may be included in the Hamiltonian of one of the leads
as: $H=H_{0}+H_{ac}=H_{0}+e\widetilde{V}\cos (2\pi ft)$, where the
unperturbed Hamiltonian, $H_{0}$, describes the leads without
microwaves. The effect of the oscillating potential is that the
time-dependent part of the electron wave function in this lead,
when expanded into a power series, contains energy components at
$E$, $E\pm hf$, $E\pm 2hf$, ..., etc. These are called sidebands.
The expansion can be done as follows \cite{Tien}:
\begin{eqnarray}
\psi (r,t) &=&\varphi (r)\exp \left( -i\int
dt[E+e\widetilde{V}\cos (2\pi
ft)]/\hbar \right)  \nonumber \\
&=&\varphi (r)\exp \left( -iEt/\hbar \right) \sum_{n=-\infty
}^{\infty
}J_{n}(e\widetilde{V}/hf)\exp (-in2\pi ft)  \nonumber \\
&=&\varphi (r)\left( \sum_{n=-\infty }^{\infty
}J_{n}(e\widetilde{V}/hf)\exp (-i[E+nhf]t/\hbar )\right)
\label{wavefunction}
\end{eqnarray}
$\varphi (r)$ is the eigenfunction satisfying $H_{0}\varphi
(r)=E\varphi (r)$ {\bf \ }and forms the spatial part of the wave
function $\psi (r,t)$. $ J_{n}(\alpha )$ is the $n$th order Bessel
function of the first kind (see Fig$.$ 1a) evaluated at $\alpha
=e\widetilde{V}/hf$.

Let $\rho _{l}$ and $\rho _{r}$ be the unperturbed densities of
states of the left and right leads. The tunnel current through a
junction without a microwave field is then given by
\begin{equation}
I(V_{SD})=c\int_{-\infty }^{\infty
}dE[f_{l}(E-eV_{SD})-f_{r}(E)]\rho _{l}(E-eV_{SD})\rho _{r}(E)
\end{equation}
where $V_{SD}$ is the source-drain voltage, $f(E)$ is the Fermi
function, and $c$ is a constant proportional to the tunnel
conductance.

\begin{figure}[htbp]
  \begin{center}
  \centerline{\epsfig{file=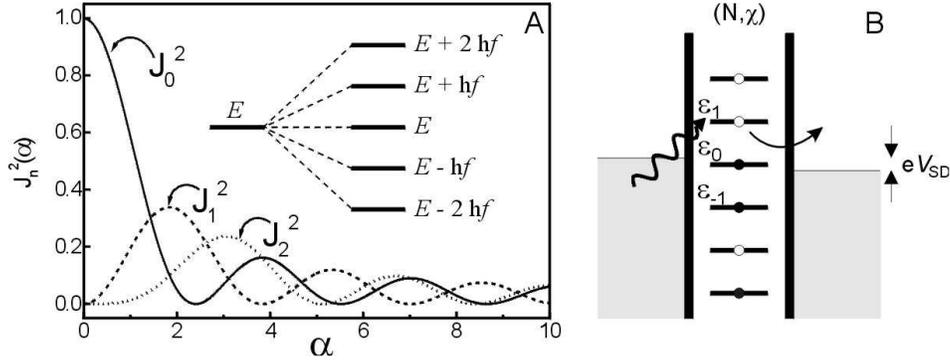, width=12.5cm, clip=true}}
    \caption{(a) Squared Bessel functions of the first kind
    $J_n^2(\alpha)$, for $n$ = 0, 1 and 2. The inset shows
    the development of sidebands of the original energy as a
    consequence of the microwave field. The population
    probability $P(n)$ of the different sidebands is given by
    $P(n)=J_n^2(e \widetilde{V}/hf)$. A positive or
    negative $n$ corresponds to the absorption or emission,
    respectively, of $n$ photons during the tunnel process.
    Elastic tunneling corresponds to $n=0$. (b) Schematic energy
    diagram of a single dot containing $N$ electrons distributed
    over the available single-particle levels $\varepsilon _j$
    in a particular configuration $\chi$. By absorbtion of a photon,
    an electron can tunnel into $\varepsilon_1$ so that the
    electron number changes into $N+1$ and the configuration into
    $\chi^{\prime}$.}
    \label{fig1}
  \end{center}
\end{figure}

From Eq$.$ (\ref{wavefunction}), we can write an effective density
of states in one of the leads (we choose the right lead) given by:
\begin{equation}
\widetilde{\rho }_{r}(E)=\sum_{n=-\infty }^{\infty }\rho
_{r}(E+nhf)J_{n}^{2}(e\widetilde{V}/hf)
\end{equation}
If tunneling is a weak perturbation, the dc current in the
presence of microwaves, $\widetilde{I\text{,}}$ is given by
\cite{Tien}:
\begin{align}
\widetilde{I}(V_{SD}) = & \hspace{0.1cm} c\sum_{n=-\infty}^{\infty}J_{n}^{2}(e\widetilde{V}/hf) \times \nonumber \\
& \int_{-\infty }^{\infty }dE[f_{l}(E-eV_{SD})-f_{r}(E+nhf)]\rho
_{l}(E-eV_{SD})\rho _{r}(E+nhf)  \nonumber \\
= & \sum_{n=-\infty }^{\infty }J_{n}^{2}(e\widetilde{V}/hf)
I(V_{SD}+nhf/e)
\end{align}
We stress that tunneling is assumed to be a weak perturbation,
implying that the sidebands are only well-defined for $f\gg \Gamma
.$ Since there is no electric field in the scattering-free leads,
mixing of electron states \cite {Platero}, or photon absorption is
absent in the leads. A positive or negative $n$ corresponds to the
absorption or emission, respectively, of $n$ photons during the
tunnel process. Elastic tunneling corresponds to $n=0$. {\it For a
single junction the dc current in the presence of microwaves is
thus described, simply in terms of the dc current without
microwaves.} From the normalization $\sum_{n=-\infty }^{\infty
}J_{n}^{2}(\alpha )=1$, it follows that the integrated current
does not change due to the oscillating potential: $\int
\widetilde{I}(V_{SD})dV_{SD}=\int I(V_{SD})dV_{SD}.$ We emphasize
that although the oscillating field is entirely classical, the
interaction with an electron, described by the Schr\"{o}dinger
equation, is only via exchange of {\it discrete }energy quanta.
Equation (4) is only valid for single junctions where the
tunneling takes place via a single hop. An extension of Eq$.$ (4)
to describe a double junction system is discussed next.

\subsection{Master equation for PAT through a quantum dot}
\label{master}

Electron transport through double barrier structures is resonant
when the Fermi energy of the leads aligns with a discrete energy
state between the two barriers. Transport through semiconductor
quantum wells are usually well described by non-interacting
electron models. Also, their transport properties in the presence
of an oscillating signal can, in a first-order approximation, be
described by the time-dependent, non-interacting Schr\"{o}dinger
equation. The result of such calculations is that, next to the
main resonance, extra peaks appear at distances corresponding to
the photon energy, $hf$ \cite{Sokolovski,Johansson}.

In quantum dots electrons are confined in all directions. The
total number of electrons, and the total charge, is thus a
discrete value. This makes it essential to include the Coulomb
interactions when describing transport. The standard model is
known as the single-electron tunneling, or the Coulomb blockade
model \cite{LesHouches}. This model takes into account that at low
voltages and low temperatures only one electron can tunnel at a
time. The necessary energy to add an extra electron to a quantum
dot consists of the charging energy $E_{c}=e^{2}/C$ for a single
electron, and a discrete energy difference, $\Delta \varepsilon $,
arising from the quantum-mechanical confinement. In practice, a
quantum dot has discrete energy states if $ \Delta \varepsilon $
exceeds the thermal energy $k_{B}T$ \cite{Coulomb}{\bf . }Assuming
sequential tunneling of single electrons, the current can be
calculated with a master equation \cite{Averin,Beenakker}.

PAT through small systems in which Coulomb blockade is important
was considered first by Likharev and Devyatov \cite{Likharev},
Hadicke and Krech \cite{Hadicke}, and Bruder and Schoeller
\cite{bruder}. A direct inclusion of the Tien and Gordon equations
\cite{Tien} in a master equation that takes into account Coulomb
blockade \cite{LesHouches} can be made by writing the tunnel rate
through each barrier in the presence of microwaves $\widetilde{
\Gamma }(E)$ in terms of the rates without microwaves $\Gamma (E)$
\cite {Leo-PAT1}:
\begin{eqnarray}
\widetilde{\Gamma
}(E)=\sum\limits_{n=-\infty}^{+\infty}J_{n}^{2}(\alpha )\Gamma
(E+nhf)
\label{PAT-rate}
\end{eqnarray}
Equation \ref{PAT-rate} has a direct link to studies of the
effects of fluctuations in the electromagnetic environment on
single-electron tunneling \cite{Ingold-Nazarov}. If the spectral
density of the environment is characterized by the probability
function $P(hf)$, then the rate including the environment $\Gamma
_{env}(E)$, can be written in terms of the rate without the
environment $\Gamma (E)$ as \cite{Hu-Connel}:
\begin{eqnarray}
\Gamma _{env}(E)= \int_{- \infty}^{\infty} d(hf)P(hf)\Gamma (E+hf)
\label{Env-rate}
\end{eqnarray}
Whereas Eq$.$ (\ref{PAT-rate}) describes a monochromatic
environment, the fluctuations in general are broad band in
frequency, as described in Eq$.$ (\ref{Env-rate}). Examples of
environments that have been studied experimentally, are the
impedance in the leads \cite{Ingold-Nazarov}, blackbody radiation
\cite{Harvard}, and phonons \cite{Toshi-Spont}. We note, however,
that Eqs$.$ (\ref{PAT-rate},\ref{Env-rate}) are valid only for
systems with a continuous density of states (i$.$e$.$ $\Delta
\varepsilon $ $\ll k_{B}T$) and immediate relaxation to the ground
state after each tunnel event.

In the case of quantum dots with large level separation (i$.$e$.$
$\Delta \varepsilon $ $\gg k_{B}T$), one needs to keep track of
the occupation probabilities of each discrete state. This
increases the amount of bookkeeping, but has the advantage that
intra-dot relaxation and excitation processes can be included. In
our model \cite{Tjerk-PAT2} for PAT through small dots, we assume
$E_{c}\gg \Delta \varepsilon $, $k_{B}T$, $eV_{SD}$, $ nhf$, such
that we only need to consider two charge states (i$.$e$.$ the
electron number is either $N$ or $N+1$) \cite{Wan}. We neglect
level broadening due to a finite lifetime of the electrons on the
dot.

A charge state (Fig$.$ 1b) is described by the electron number
$N$, together with the particular occupation of the electrons in
the available single-particle levels $\{\varepsilon _{j}\}$. If
$N$\ electrons are distributed over $k$ levels, the number of
distinct dot configurations, $\chi ,$ is given by $\binom{k}{N}$.
The probability, $P_{N,\chi }$, for state $(N,\chi ) $ is
calculated from a set of master equations given by:
\begin{eqnarray}
\dot{P}_{N,\chi } &=&\sum_{\chi ^{^{\prime }}}P_{N+1,\chi ^{\prime
}}(\Gamma _{l,j_{\chi ^{\prime }}}^{out}+\Gamma _{r,j_{\chi
^{\prime }}}^{out})
\nonumber \\
&&-P_{N,\chi }\sum_{j=empty}(\Gamma _{l,j}^{in}+\Gamma
_{r,j}^{in})
\label{rate equation} \\
&&+\sum_{\chi ^{\prime \prime }\neq \chi }P_{N,\chi ^{\prime
\prime }}\Gamma _{\chi ^{\prime \prime }\rightarrow \chi
}-P_{N,\chi }\sum_{\chi ^{\prime \prime \prime }\neq \chi }\Gamma
_{\chi \rightarrow \chi ^{\prime \prime \prime }}  \nonumber
\end{eqnarray}
and the equivalent forms for $\dot{P}_{N+1,\chi ^{\prime }}$. To
find a stationary solution, these equations are all set to zero
($\dot{P}=0$) and solved with the boundary condition:
\begin{equation}
\sum_{\chi }P_{N,\chi }+\sum_{\chi ^{\prime }}P_{N+1,\chi ^{\prime
}}=1 \label{boundary condition}
\end{equation}
For $N=2$ distributed over five different single-particle levels
$\{\varepsilon _{j}\}$, there are ten different configurations,
$\chi $, yielding ten equations for $\dot{P}_{N,\chi }$ and also
ten equations for $\dot{P}_{N+1,\chi ^{\prime }}$.

The first and second term in Eq$.$ (\ref{rate equation})
correspond to a change in the occupation probability of a certain
distribution due to tunneling (the number of electrons on the dot
changes). In the first term an electron tunnels out of the dot.
Only those rates are taken into account that correspond to an
electron tunneling out of state $j_{\chi ^{\prime }}$ that leave
the dot in the distribution ($N$, $\chi $). In the second term, an
electron tunnels onto the dot. One needs to sum over all the
states $j$ that are empty when the dot is in configuration $\chi
$, because all these events cause a transition from state $(N,\chi
)$ to a state $(N+1,\chi ^{\prime })$. $\Gamma _{l,j}^{in}\ $and
$\Gamma _{l,j}^{out}$ are the tunnel rates through the left
barrier in and out of single-particle level $j$ on the dot:
\begin{eqnarray}
\Gamma _{l/r,j}^{in}(\varepsilon _{j}) &=&\Gamma
_{l/r,j}\sum_{n}J_{n}^{2}(\alpha _{l/r})f(\varepsilon
_{j}-\frac{C_g}{C}eV_{g}-nhf+\eta _{l/r}eV_{SD};T_{l/r})  \label{rates} \\
\Gamma _{l/r,j}^{out}(\varepsilon _{j}) &=&\Gamma
_{l/r,j}\sum_{n}J_{n}^{2}(\alpha _{l/r})[1-f(\varepsilon
_{j}-\frac{C_g}{C}eV_{g}-nhf+\eta _{l/r}eV_{SD};T_{l/r})]
\nonumber
\end{eqnarray}
where $\Gamma _{l/r,j}$ is the tunnel rate through the left or
right barrier of energy level $j$. $\alpha _{l/r}$ is the
parameter describing the microwave field at the left or right
barrier, $C_{g}$ is the gate capacitance, $C$ is the total dot
capacitance, $T_{l/r}$ is the temperature of the left or right
lead. $\eta _{l/r}$ is a parameter describing the asymmetry of the
dc voltage drop across the two barriers. We assume that the tunnel
rates, $\Gamma _{l/r,j}$, do depend on the level index $j$, but
that they are independent of energy.

In the last two terms of Eq$.$ (\ref{rate equation}), the number
of electrons on the dot is fixed, while only the distribution of
the electrons over the states changes. This includes effects from
relaxation (i$.$e$.$ intra-dot transitions, $\chi ^{\prime \prime
}\rightarrow \chi $ and $\chi \rightarrow \chi ^{\prime \prime
\prime },$ where the total energy decreases) or excitation inside
the dot (i$.$e$.$ intra-dot transitions, $\chi ^{\prime \prime
}\rightarrow \chi $ and $\chi \rightarrow \chi ^{\prime \prime
\prime },$ where the total energy increases). Below, we take
excitation rates equal to zero (i$.$e$.$ no intra-dot absorption)
but, allow for non-zero relaxation rates.

An expression for the dc current can be found by calculating the
net tunnel rate through one of the barriers. Using the
probabilities $P_{N,\chi }$ and the tunnel rates through the left
barrier, this leads to:
\begin{equation}
I=e\sum_{\chi }\sum_{j=empty}P_{N,\chi }\Gamma
_{l,j}^{in}-e\sum_{\chi ^{\prime }}\sum_{j=full}P_{N+1,\chi
^{\prime }}\Gamma _{l,j}^{out}
\end{equation}
{\bf \ }

In the numerical calculations in the next section we take equal ac
amplitudes dropping across the left and right barriers; i$.$e$.$
$\alpha _{l}=\alpha _{r}=\alpha $.

\subsection{Numerical results}
\label{numresults}

\begin{figure}[htbp]
  \begin{center}
  \centerline{\epsfig{file=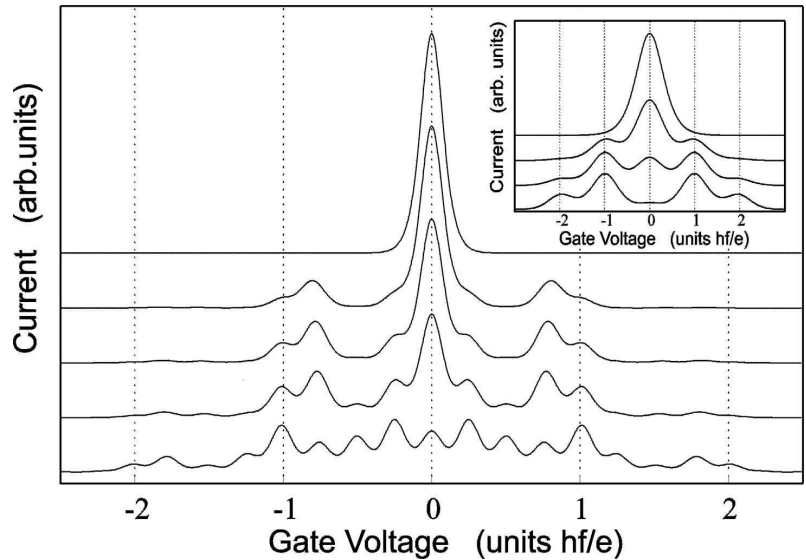, width=10.5cm, clip=true}}
    \caption{Calculation without relaxation. Curves are offset for clarity. The parameters for the
    data in the inset are $\Delta \varepsilon =3hf$, $hf=5k_{B}T$, and
    from top to bottom $\alpha =0$, 1, 1.5, 2. The parameters for the
    main figure are $ \Delta \varepsilon =0.75hf$, $hf=20k_BT$\ and
    from top to bottom $\alpha =0 $, 0.5, 0.75, 1, 1.5.}
    \label{fig2}
  \end{center}
\end{figure}

Figure \ref{fig2} shows calculations without relaxation between
the states in the dot. The inset shows the case for transport
through only a single level; $\Delta \varepsilon \gg hf$. Next to
the main resonance, side-peaks develop at multiples of $hf/e$ when
the microwave power is increased via the parameter $\alpha .$ The
broadening of the resonances is due to a finite temperature. In
the main figure transport can also occur via excited states; here
$\Delta \varepsilon < hf$. Not only side-peaks develop, but also
peaks at other gate voltages. These peaks arise due to the
interplay between the discrete single-particle states and the
photon energy. Their locations are given by $(m\Delta \varepsilon
+nhf)/e$ where $m=0,$ $ \pm 1,$ $\pm 2,$ ... and $n$ is the photon
number. Similar simulation results have been reported by Bruder
and Schoeller \cite{bruder}.

\begin{figure}[htbp]
  \begin{center}
  \centerline{\epsfig{file=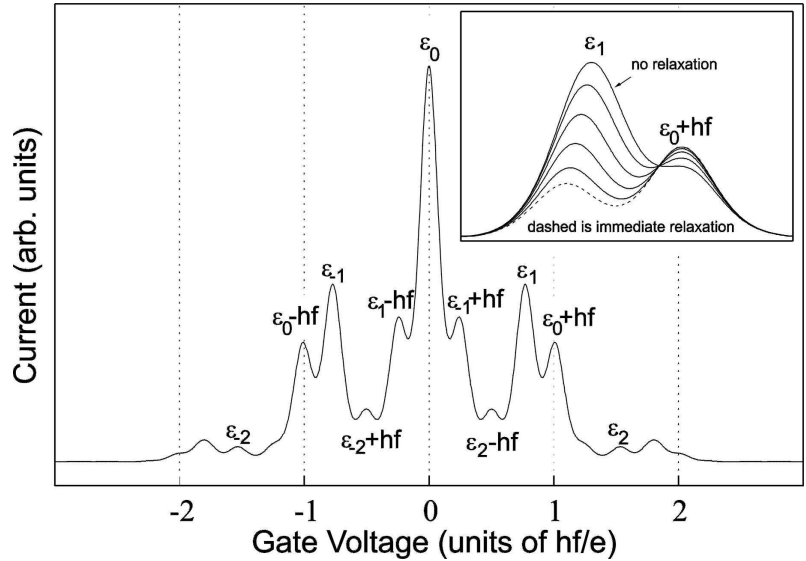, width=10.5cm, clip=true}}
    \caption{Expansion for the curve $\alpha =1$ from Fig$.$ 2. The inset
    shows the effect of an increasing relaxation rate. The relaxation
    rates divided by the tunnel rate are 0, 0.1, 0.35, 1, 3.5, and
    infinite.}
    \label{fig3}
  \end{center}
\end{figure}

Figure \ref{fig3} shows an expansion for the curve with $\alpha
=1$. We have assigned the excited states and the particular PAT
processes. The highest occupied single-particle level of the $N+1$
ground state is denoted by $\varepsilon _{j}$ with $j=0$; positive
$j$'s are excited states above $ \varepsilon _{0}$ and negative
$j$'s are below $\varepsilon _{0}$ (see Fig$.$ 1b). The inset
shows the effect of relaxation. Upon increasing the relaxation
rate, the peaks corresponding to tunneling through excited states
decrease, while peaks increase when tunneling occurs through the
ground state.
\begin{figure}[htbp]
  \begin{center}
  \centerline{\epsfig{file=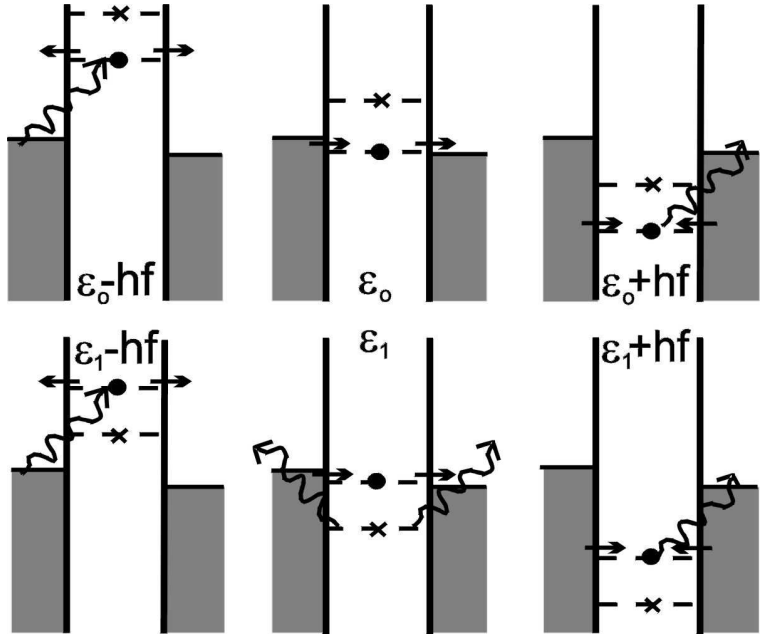, width=10cm, clip=true}}
    \caption{Diagrams depicting the tunneling events which dominantly contribute to the current
through a quantum dot at different gate voltages. A small dc bias
raises the left Fermi level with respect to the right Fermi level.
$\varepsilon _{0}$ and $\varepsilon _{1}$ denote the ground state
and the first excited state of the $(N+1)$ electron system. When
the $(N+1)^{th}$ electron tunnels to one of the two reservoirs,
the energy states of the dot drop by the charging energy $ E_{c}$.
The corresponding diagrams for $N$ electrons are not shown.}
    \label{fig4}
  \end{center}
\end{figure}

To explain these numerical results, we show energy diagrams in
Fig$.$ \ref{fig4}, assuming that only the highest two
single-particle levels contribute to the current for the
transition between $N$ and $N+1$ electrons on the dot. For small
dc bias voltage and no ac voltages a current resonance occurs when
the topmost energy state (i$.$e$.$ the electrochemical potential)
of the quantum dot lines up with the Fermi levels of the leads
(see the diagram $\varepsilon _{0}$). When high-frequency voltages
drop across the two barriers, additional current peaks appear. We
distinguish two mechanisms. The first mechanism gives photon
induced current peaks when the {\it separation} between the ground
state $\varepsilon _{0}$ and the Fermi levels of the leads {\it
matches} the photon energy (or multiples, $nhf$), as depicted in
the diagrams labeled by $\varepsilon _{0}+hf$ and $\varepsilon
_{0}-hf$. The minus and plus signs correspond to being before or
beyond the main resonance. Note that also the case of $\varepsilon
_{0}-hf$ involves photon absorption. Following the literature on
the tunneling time, we call these current peaks: {\it sidebands}
\cite{tunnel-time}. The second mechanism leads to photon peaks
when an excited state is in resonance with the Fermi levels of the
leads (see diagram $\varepsilon _{1}$). Without PAT, transport
through the excited state, $\varepsilon _{1}$, is blocked since
Coulomb blockade prevents having electrons in both the ground
state and the excited state simultaneously. The electron in the
ground state cannot escape from the dot, because its energy is
lower than the Fermi levels in the leads. PAT, however, can empty
the ground state $\varepsilon _{0}$ when the electron absorbs
enough energy and leaves the dot. This process is analogous to
photo-ionization. Now, the ($N+1)^{\text{th}}$ electron can tunnel
resonantly via the excited state $\varepsilon _{1}$ as long as the
state $ \varepsilon _{0}$ stays empty. Note that for this second
mechanism $nhf$ has to {\it exceed}, but not necessarily {\it
match} the energy splitting $ \Delta \varepsilon =\varepsilon
_{1}-\varepsilon _{0}$. It is clear from these diagrams that
relaxation from $\varepsilon _{1}$ to $\varepsilon _{0}$ decreases
the height of this resonant peak. More photon peaks are generated
when these two mechanisms are combined as in the diagrams labeled
by $ \varepsilon _{1}+hf$ and $\varepsilon _{1}-hf$. We thus see
that PAT can populate the excited states by tunneling between dot
and leads. So, even without intra-dot transitions, we can perform
photon spectroscopy on discrete quantum dot states.

\subsection{Photon-assisted pumping}
\label{PAP}

It is important to note that in the diagrams of Fig$.$ \ref{fig4}
only processes with tunneling from or to states in the leads close
to the Fermi levels contribute to the net current. Tunnel
processes that start with an electron in one of the leads from
further below the Fermi level are cancelled by an electron from
\begin{figure}[htbp]
  \begin{center}
  \centerline{\epsfig{file=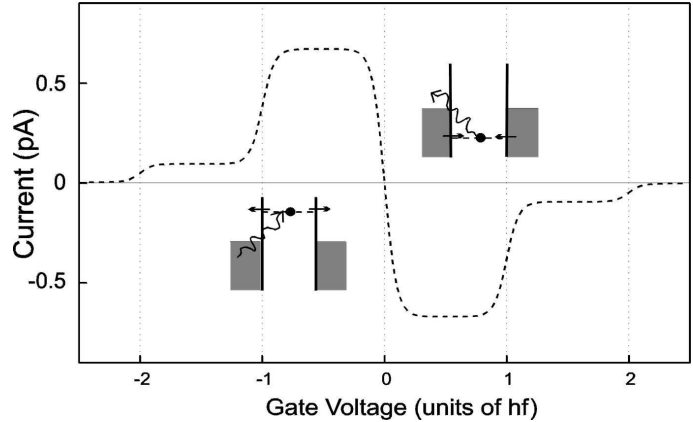, width=9cm, clip=true}}
    \caption{Calculation of the
current for zero-bias voltage $(V=0)$\ as a function of gate
voltage in the case where the ac voltage drop over one barrier is
5\% smaller than over the other barrier $(k_{B}T=0.05hf$, $\Gamma
=5 10^{8}$\ s$^{-1})$. The insets depict which tunneling events
are responsible for the pumped current when the ground state of
the quantum dot is below or above the Fermi levels of the leads.}
    \label{fig5}
  \end{center}
\end{figure}
the other lead. However, this is only true when the ac voltage
drop is the same for both barriers. When the ac voltage drops
across the two barriers are unequal, the dot acts as an electron
pump \cite{Leopump,Leo-PAT1,bruder}. The resulting pumped current
makes the resonances discussed above less clear. For this reason
we discuss this pumping mechanism in more detail here before
proceeding further. Figure \ref{fig5} shows a calculation of the
pumped current as a function of the gate voltage that occurs when
the ac voltage drop over one barrier is 5\% smaller than over the
other barrier. We have taken zero dc bias voltage. To illustrate
the origin of the pumped current, the insets show the extreme
case, when all the ac voltage drop is across the left barrier. In
this case photon absorption occurs only at the left barrier. At
negative gate voltage, when the ground-state level of the dot is
{\it above} the Fermi level of the leads, an electron can {\it
enter} the dot from the left lead only (bottom left inset to
Fig$.$ \ref{fig5}). Once the electron is in the dot, it can tunnel
out through both tunnel barriers. Only tunneling to the right lead
contributes to the net current. Therefore, the net (particle)
current is to the right. When the ground-state level of the dot is
{\it below} the Fermi level of the leads, however, an electron can
only {\it leave} the dot to the left lead (upper right inset to
Fig$.$ \ref{fig5}). The dot can be filled from either lead once it
is emptied. This time only the electron tunneling {\it in} from
the right lead contributes to the net current. Therefore, there is
a net current to the left. The difference between these two
situations is the shift in the ground-state energy with respect to
the Fermi levels of the leads. So, when the gate voltage is swept
such that the ground state moves through these Fermi levels, the
pumped current changes sign. The pumped current occurs over a
width corresponding to the photon energy. The extra shoulders at
the far left and far right are due to two-photon processes.

\begin{figure}[htbp]
  \begin{center}
  \centerline{\epsfig{file=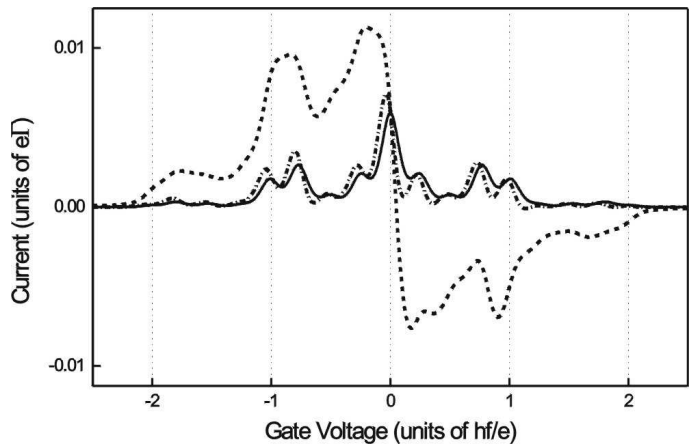, width=9cm, clip=true}}
    \caption{Simulation of the
effects of both asymmetric heating and quantum rectification on
the curve with $\alpha =1$ from Fig$.$ 3. The solid line
represents the unperturbed curve. The dashed-dotted line shows the
same curve for $T_{R}/T_{L}$ = 0.95. The dashed line shows the
influence of $\alpha _{R}/\alpha _{L}$ = 0.95 on the unperturbed
curve.}
    \label{fig6}
  \end{center}
\end{figure}

Asymmetric heating may induce a difference in the temperatures
$T_{R}$ and $ T_{L}$ of the two leads. This can also result in a
finite transport current. The effects of asymmetric heating and
asymmetric coupling of the microwave signal when a finite $V_{SD}$
is applied across the sample, is illustrated in Fig$.$ \ref{fig6}.
The solid line is a reproduction of Fig$.$ \ref{fig3}. The
dashed-dotted line shows the same trace for $T_{R}/T_{L}$ = 0.95.
The dashed line shows the influence of $\alpha _{R}/\alpha
_{L}=0.95$. It can be concluded that both effects can severely
distort the data. At small $V_{SD}$ the resonant peaks scale
linearly with $V_{SD}$, while the pumped current does not change.
Increasing the bias, while remaining in the linear regime,
therefore improves the visibility of the resonant peaks.

\section{Experiments}
\label{experiments}

\subsection{Sample geometry}
\label{sample}

\begin{figure}[htbp]
  \begin{center}
  \centerline{\epsfig{file=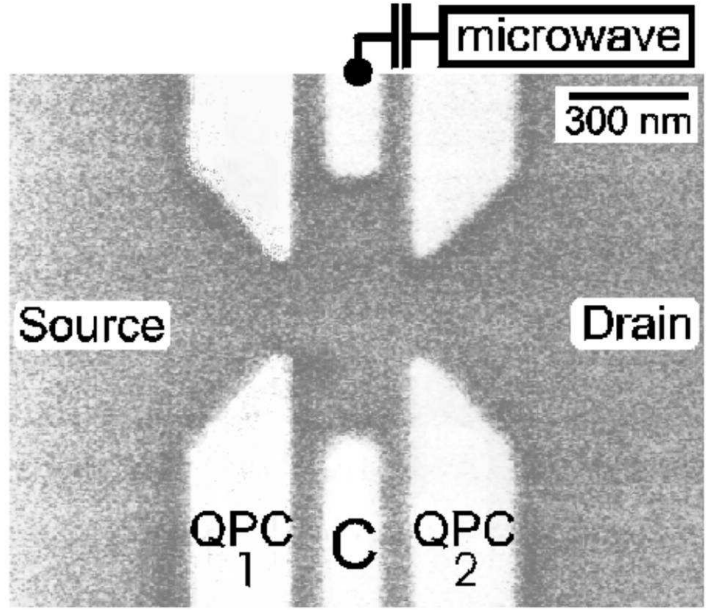, width=9.5cm, clip=true}}
    \caption{SEM photo of the
sample. The lithographic size of the dot is ($600\times 300$)
nm$^{2}$. Current can flow when we apply a voltage between source
and drain. The microwave signal is capacitively coupled to one of
the center gates.}
    \label{fig7}
  \end{center}
\end{figure}

Our measurements are performed on a quantum dot defined by
metallic gates (see Fig$.$ \ref{fig7}) in a GaAs/AlGaAs
heterostructure containing a 2-dimensional electron gas (2DEG) 100
nm below the surface. The 2DEG has mobility 2.3 10$^{6}$
cm$^{2}$/Vs and electron density 1.9 10$^{15}$ m$^{-2}$ at 4.2 K.
By applying negative voltages to the two outer pairs of gates, we
form two quantum point contacts (QPCs). An additional pair of
center gates between the QPCs confines the electron gas to a small
dot. No electron transport is possible through the narrow channels
between the center gates and the gates forming the QPCs. The
center gate voltage, $V_{g}$, can shift the states in the dot with
respect to the Fermi levels of the leads and thereby controls the
number of electrons in the dot. The energy shift is given by
$\Delta E=\kappa \Delta V_{g}$, with $\kappa$ defined as the ratio
between the dot-gate capacitance and the total capacitance of the
dot \cite{Coulomb}. A small dc voltage bias is applied between
source and drain and the resulting dc source-drain current is
measured. From standard dc measurements we find that the effective
electron temperature is approximately $T$ = 200 mK and the
charging energy $E_{c}=1$.2$ \pm $0.1 meV. We independently
determine the level splitting, $\Delta \varepsilon$, for different
magnetic fields from current-voltage characteristics. In addition
to the dc gate voltages, we couple a microwave signal (10-75 GHz)
capacitively into one of the center gates. The microwave does not
equally couple to the dot as to the leads, which results in an ac
voltage drop over both barriers.

\subsection{Frequency sensitivity of pumping}
\label{frequency}

\begin{figure}[htbp]
  \begin{center}
  \centerline{\epsfig{file=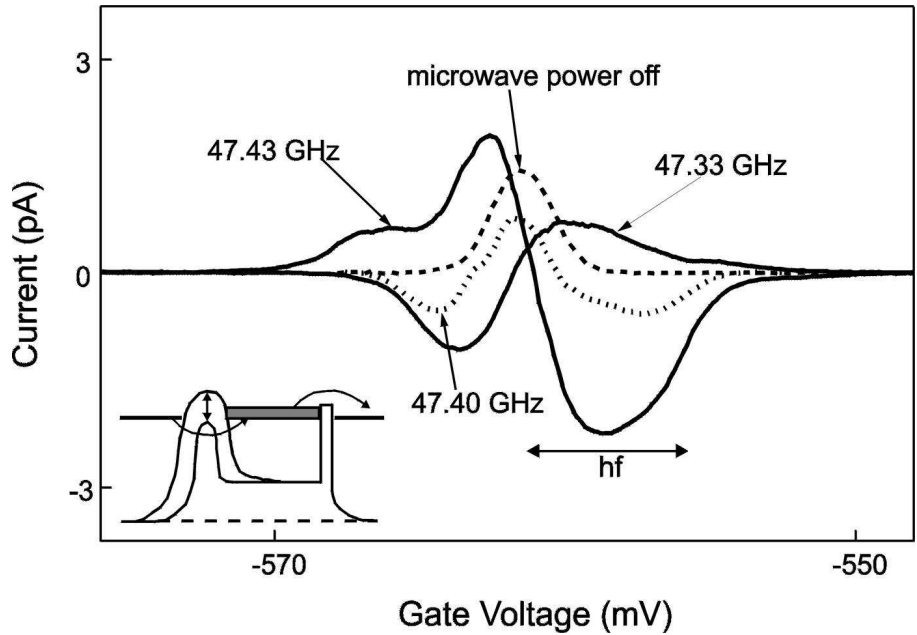, width=12cm, clip=true}}
    \caption{Measurements of the
    pumped current at $B$ = 1.96 T, $V_{SD}$ = 13 $\mu$V and
    frequencies are around 47.4 GHz. Dashed line is without
    microwaves. The dotted line shows the smallest asymmetry, but
    shows evidence for a pumping mechanism which is not included in
    our model.}
    \label{fig8}
  \end{center}
\end{figure}

We first present experimental results with a strongly pumped
current taken at $B$ = 1.96 T for three frequencies around 47.4
GHz (the arrow denotes $hf$). The dashed line in Fig$.$ \ref{fig8}
is the current without microwaves. For the lowest frequency the
current is pumped in one direction, whereas for the highest
frequency it is pumped in the opposite direction. At 47.33 GHz the
left barrier apparently has the {\it smaller} ac voltage drop,
while at 47.43 GHz the left barrier has the {\it larger} ac
voltage drop. This illustrates that the asymmetry of the voltage
drops over the two barriers sensitively depends on frequency. This
sensitivity is ascribed to standing waves in the sample holder.
The dotted line shows the current measured at an intermediate
frequency, where we expect the ac voltage drop to be equal over
both barriers. In contrast to the two solid curves, the dotted
line is lower than the dashed line without microwaves over the
whole gate voltage range. This cannot be explained by the pumping
mechanism in our model. Our model only includes the oscillation of
the potential of the leads relative to the dot, which always
results in a pumped current which changes sign at the resonance. A
negative pumped current over the whole gate voltage range, is
attributed to the effect of the microwaves on the barrier height.
The inset shows how a quantum dot can act as a pump when one
tunnel barrier is periodically modulated in height. During one
part of the cycle, when the left barrier is low, electrons enter
the dot ($\Gamma _{L}^{low}>\Gamma _{R}$ ) while they escape the
dot through the right barrier in the second half of the cycle when
the left barrier is high ($\Gamma _{L}^{high}<\Gamma _{R}$). This
is essentially a classical mechanism that has been verified
experimentally in the MHz regime \cite{Leopump}. For observing
clearly separated PAT sidebands, we first minimize pumping. For
this we measure traces of current at zero bias voltage across a
single Coulomb peak for slightly different frequencies. We finally
choose those frequency values for which the pumped current is very
small.

\subsection{PAT: Low frequency regime}
\label{PATlowfreq}

\begin{figure}[htbp]
  \begin{center}
  \centerline{\epsfig{file=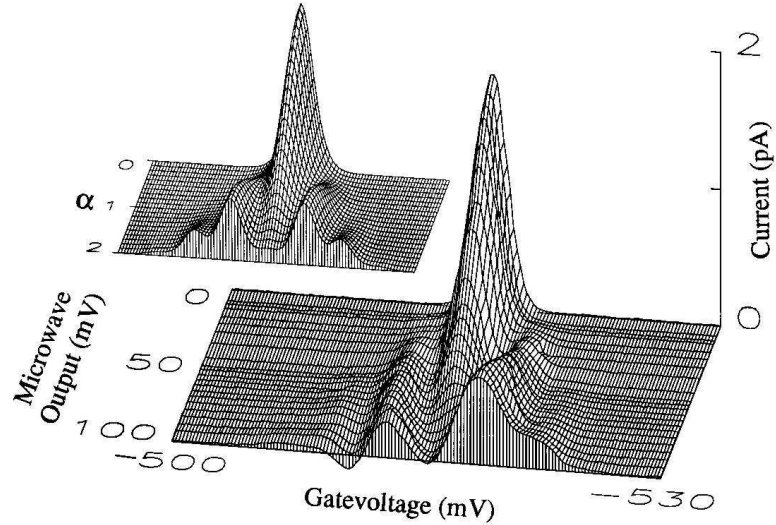, width=11cm, clip=true}}
    \caption{Measurement of the
    current through the quantum dot as a function of the center gate
    voltage and the output voltage of the microwave supply. These data
    are taken in the single-level regime ($hf < \Delta \varepsilon$).
    $hf$ = 110 $\mu$eV for $f$ = 27 GHz, $\Delta \varepsilon$ = 165
    $\mu$eV at $B$ = 0.84 T, and $V_{SD}$ = 13 $\mu$V. Inset:
    calculation of the current as a function of the gate voltage and
    the ac voltage parameter $\alpha =e\tilde{V}/hf$, taking the same
    values for $T$, $f$, and $V$ as in the experiment.}
    \label{fig9}
  \end{center}
\end{figure}

First, we study the photon sidebands of the ground state at
$B=0$.84 T \cite {B not zero}. The main part of Fig$.$ \ref{fig9}
shows measured curves of the current as a function of the gate
voltage at different microwave powers for the case $hf<\Delta
\varepsilon $. Here, current flows primarily via the ground state
and its photon sidebands (i$.$e$.$ upper diagrams in Fig$.$
\ref{fig4}). On increasing the microwave power, we see that the
height of the main resonance decreases to zero while additional
resonances develop with increasing amplitude. When we convert gate
voltage to energy, we find that the additional resonances are
located at $\varepsilon _{0}\pm hf$ and $ \varepsilon _{0}\pm 2hf$
\cite{2hf>Deltae}. The power dependence is in agreement with the
behavior of the Bessel functions: $J_{0}^{2}(\alpha )$ for the
main resonance $\varepsilon _{0}$, $J_{1}^{2}(\alpha )$ for the
one-photon sidebands $\varepsilon _{0}\pm hf$, and
$J_{2}^{2}(\alpha )$ for the two-photon sidebands $\varepsilon
_{0}\pm 2hf$. For comparison, we show a calculation in the inset
to Fig$.$ \ref{fig9} for the same values for the temperature,
frequency and bias voltage as in the experiment. We have assumed
equal ac voltages across the two barriers. The difference between
measured and calculated data is attributed to an asymmetry in the
ac coupling.
\subsection{PAT: High frequency regime}
\label{PAThighfreq}

\begin{figure}[htbp]
  \begin{center}
  \centerline{\epsfig{file=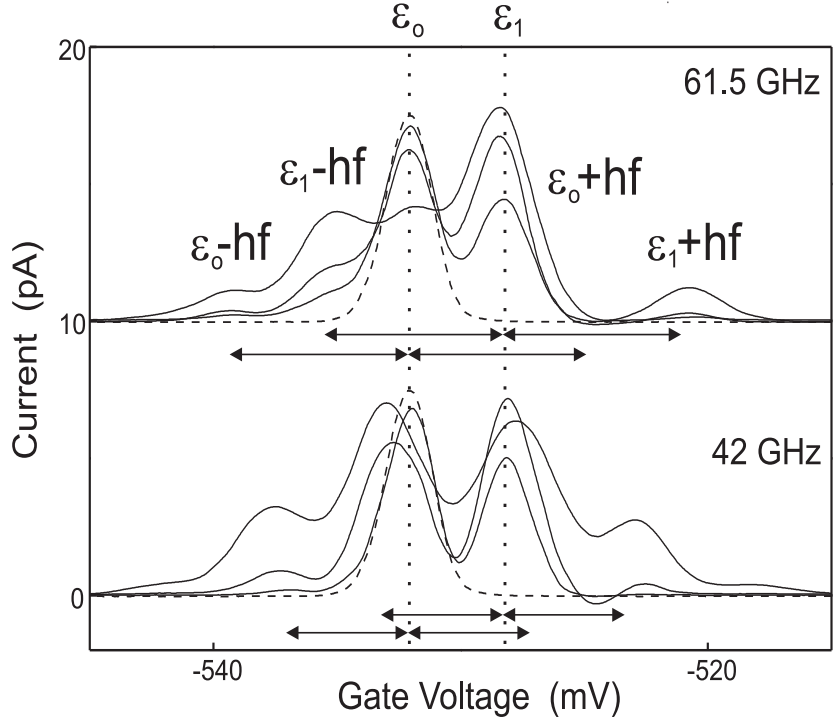, width=11cm, clip=true}}
    \caption{Measured current as a
    function of center gate voltage for different microwave powers.
    The dashed curve is without microwaves. $B$ = 0.91 T, $V_{SD}$ = 13
    $\mu$V. $f$ = 61.5 GHz in the top section, $f$ = 42 GHz in the
    bottom section. As the frequency is reduced between top and bottom
    sections, the ground-state resonance $\varepsilon _0$ and the
    resonance attributed to the excited state $\varepsilon _1$
    remain at the same gate voltage position. The other peaks,
    $\varepsilon _0-hf$ and $\varepsilon_1 \pm hf$, shift inward
    by an amount which corresponds to the change in photon energy as
    indicated by the arrows. We do
    not observe $\varepsilon _{0}+hf$\ in this measurement.}
    \label{fig10}
  \end{center}
\end{figure}

\begin{figure}[htbp]
  \begin{center}
  \centerline{\epsfig{file=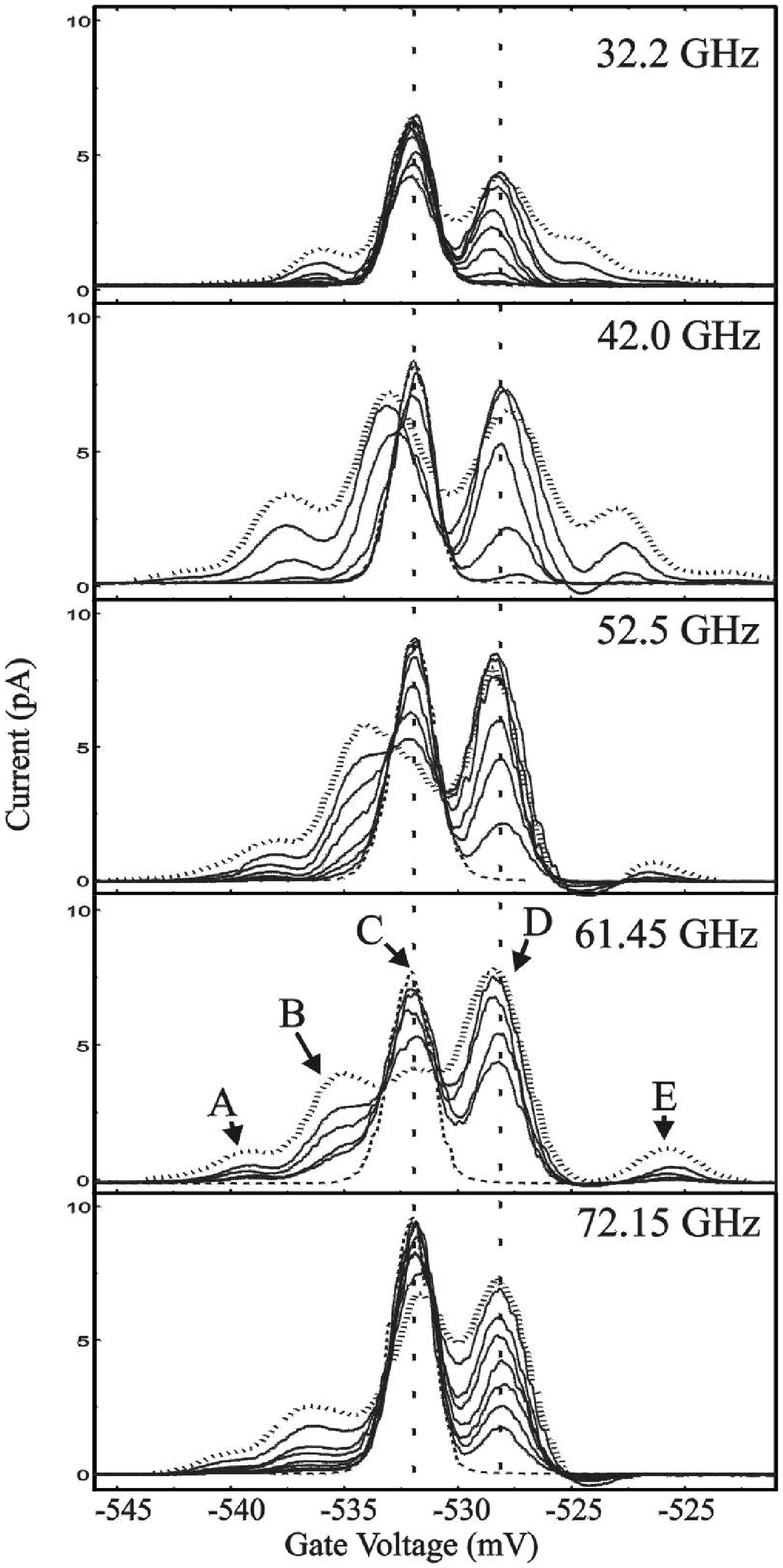, width=8cm, clip=true}}
    \caption{$I_{DC}-V_{g}$
    curves for increasing frequency; $B$ = 0.91 T, $V_{SD}$ = 13 $\mu$V
    and powers range between -30 and 0 dBm. The dashed curve is
    without microwave power. The dotted curve is with maximum power.
    Five different peaks can be distinguished in the data. These
    peaks, labelled A through E for the 61.45
    GHz traces, correspond to the $\varepsilon_0 - hf$, $\varepsilon_1 - hf$,
    $\varepsilon_0$, $\varepsilon_1$ and $\varepsilon_{1}+hf$
    peaks, respectively.}
    \label{fig11}
  \end{center}
\end{figure}

We now discuss the higher frequency regime where $hf>\Delta
\varepsilon$, such that PAT can induce current through excited
states. Figure \ref{fig10} shows the current at $B$ = 0.91 T (here
$\Delta \varepsilon =130$ $\mu $eV). In the top section $f$ = 61.5
GHz ($hf$ = 250 $\mu$eV) and in the bottom section $f$ = 42 GHz
($hf$ = 170 $\mu$eV). As we increase the power, we see extra peaks
coming up. We label the peaks as in Fig$.$ \ref{fig4}. On the
right side of the main resonance a new peak appears, which we
assign to photo-ionization, followed by tunneling through the
first excited state. At higher powers the one-photon sidebands of
the main resonance as well as those of the excited state resonance
appear. We do not observe the peak for $\varepsilon_{0}+hf$, in
this measurement. This can be explained, at least in part, by the
fact that here an electron can also tunnel into $\varepsilon
_{1}$, which blocks the photon current through $ \varepsilon
_{0}+hf$. Simulations confirm that the peak for $\varepsilon
_{0}+hf$ can be several times weaker than the peak for
$\varepsilon _{0}-hf$ \cite{Tjerk-PAT2}. Also, it is masked by the
high peak for $\varepsilon _{1}$ right next to it. The arrows
underneath the curves mark the photon energy. The peaks
$\varepsilon _{0}$ and $\varepsilon _{1}$ remain in place when we
change the frequency, since the photon energy evidently does not
alter the energy splitting. The other peaks, $\varepsilon _{0}-hf$
and $\varepsilon _{1}\pm hf$, shift by an amount that corresponds
to the change in photon energy as indicated by the arrows. This
reflects that the sidebands originate from matching the states
$\varepsilon _{0}$ and $\varepsilon _{1}$ to the Fermi levels of
the leads by a photon energy $hf$. Figure \ref{fig11} shows a
large data set. In each panel, different traces are taken at
different microwave power. The panels differ in frequency. We
further substantiate the peak assignment below by studying
detailed frequency, magnetic field and power dependence.

\subsection{Frequency dependence}
\label{freqdep}

\begin{figure}[htbp]
  \begin{center}
  \centerline{\epsfig{file=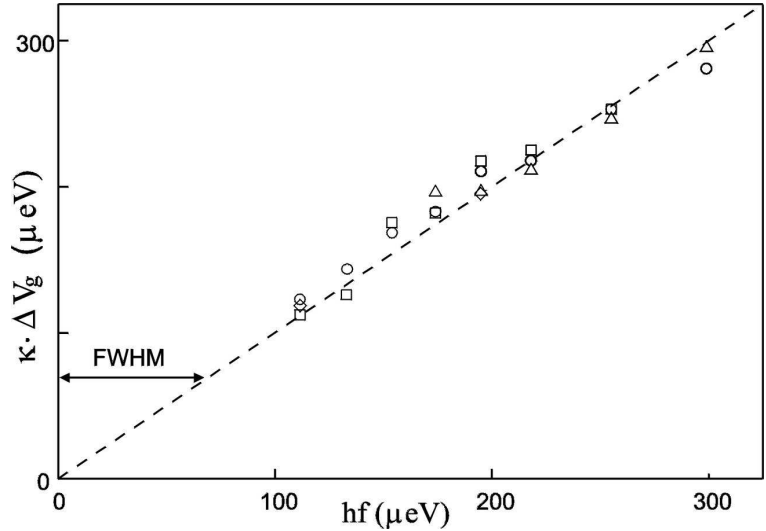, width=10cm, clip=true}}
    \caption{Peak spacings versus
    photon energy. $\square$: spacing between $\varepsilon_{0}$ and
    $\varepsilon _{0}-hf $. $\lozenge$: spacing between $\varepsilon_{0}$
    and $\varepsilon _{0}+hf$. $\triangle $: spacing between
    $\varepsilon_1$ and $\varepsilon_1-hf$. {\large $\circ$}:
    spacing between $\varepsilon_1$ and $\varepsilon_1+hf$.
    The dashed line is based on the gate voltage to energy conversion
    factor $\kappa$ determined independently from dc measurements,
    and has the theoretically expected slope equal to 1. The arrow
    indicates the FWHM of the main resonance.}
    \label{fig12}
  \end{center}
\end{figure}

Figure \ref{fig12} shows the spacing between a resonance and its
photon sidebands as a function of the photon energy. Different
markers correspond to different photon sidebands. The factor
$\kappa =3$5 $\mu eV/mV$, to convert the peak spacings in mV gate
voltage into energy, is determined from dc measurements. The full
width at half maximum (FWHM) of the resonance without microwaves,
indicated by the arrow, is proportional to the effective electron
temperature in the leads. Structure due to photon energies below
this value is washed out by the thermal energy $k_{B}T$. The
frequency scaling firmly establishes PAT as the transport
mechanism [9,11-14]. 
The observation that the sidebands move linearly with frequency,
while the ground and excited state resonances stay fixed, supports
our identification of the different peaks.

\subsection{Magnetic field dependence}
\label{Bdep}

\begin{figure}[htbp]
  \begin{center}
  \centerline{\epsfig{file=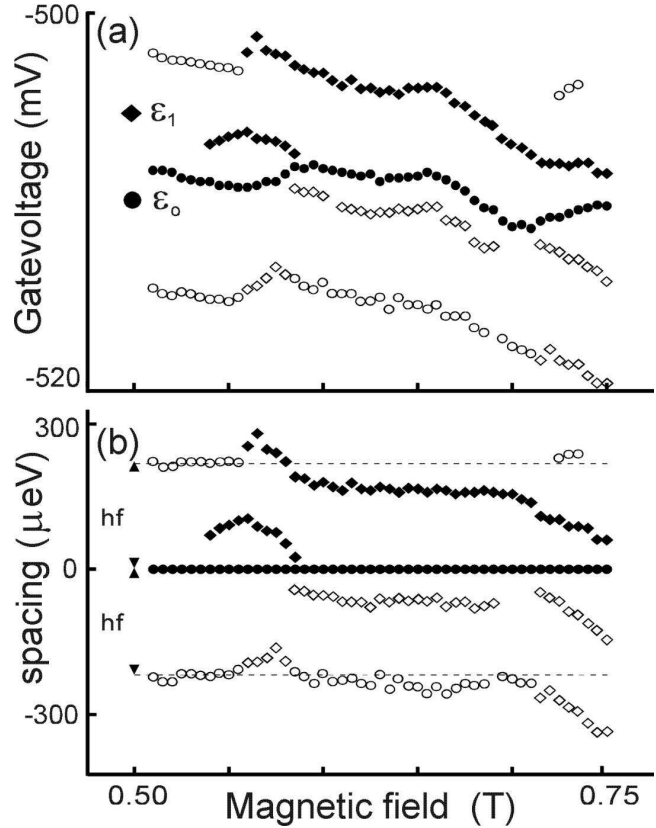, width=8.5cm, clip=true}}
    \caption{(a) Peak positions in
    gate voltage versus magnetic field at 52.5 GHz. Solid
    symbols denote peaks which are independent of frequency. Open
    symbols denote peaks that scale with frequency. (b) Peak spacings
    relative to the main resonance converted to energy. Closed
    circles: $\varepsilon_{0}$; open circles: $\varepsilon_{0} \pm
    hf$; closed diamonds: $\varepsilon_{1}$; open diamonds:
    $\varepsilon_{1}-hf$\ and $\varepsilon _{1}-2hf$.}
    \label{fig13}
  \end{center}
\end{figure}

We now use a magnetic field to change the energy separation
between the ground state and the first excited state
\cite{Coulomb,Johnson}, while keeping the distance to the
sidebands fixed. Figure \ref{fig13}a shows the positions in gate
voltage of all observed peaks for 52.5 GHz as a function of
magnetic field. The filled circles reflect the evolution of
$\varepsilon _{0}$ with magnetic field. This ground state weakly
oscillates with a periodicity of $\sim $80 mT which roughly
corresponds to the addition of an extra flux quantum to the dot.
The filled diamonds reflect the evolution of $ \varepsilon _{1}$.
The open circles (diamonds) show the sidebands $ \varepsilon
_{0}\pm hf$ ($\varepsilon _{1}\pm hf$). Figure \ref{fig13}b shows
the magnetic field evolution of the excited state and the photon
sideband peaks relative to the ground state (i$.$e$.$ we have
subtracted $ \varepsilon _{0}(B)$ from the other curves). We see
that the energy splitting decreases on increasing the magnetic
field and for $0.54$ T $ <B<0.58$ T a degeneracy of the ground
state is temporarily lifted and actually two excited states are
observed \cite{fielddependence}. The dashed lines denote the
photon energy $hf$ = 217 $\mu$eV for 52.5 GHz. The open circles
close to these lines are the photon processes $\varepsilon_{0}\pm
hf $, demonstrating that they indeed move together with the ground
state. The open diamonds are the $\varepsilon_{1}-hf$ and
$\varepsilon _{1}-2hf$ processes. Their motion follows the motion
of $\varepsilon _{1}$. We have thus shown that we can vary the
states $\varepsilon _{0}$ and $\varepsilon _{1}$ with the magnetic
field and, independently, vary the separation to the sidebands
with the microwave frequency.

\subsection{Power dependence}
\label{Pdep}

\begin{figure}[htbp]
  \begin{center}
  \centerline{\epsfig{file=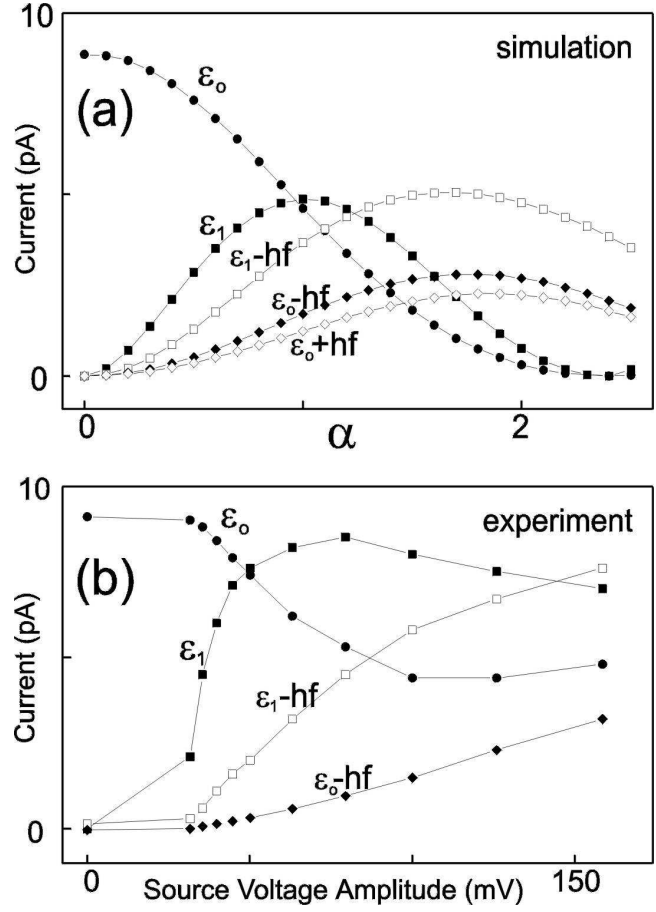, width=8.5cm, clip=true}}
    \caption{(a) Calculation of the
    peak heights as a function of the ac voltage drop across the
    barriers $\alpha =\frac{e\widetilde{V}}{hf}$ ($T$ = 200 mK, $V$ = 13
    $\mu$V, $f$ = 52.5 GHz). The tunnel rates from the leads to the
    ground state and the excited state are set to
    $\Gamma_{\varepsilon_0}$ = 5 10$^{8}$ s$^{-1}$ and
    $\Gamma_{\varepsilon_1}$ = 14 10$^{8}$ s$^{-1}$, respectively. The
    relaxation rate from the excited state to the ground state is
    assumed to be zero in the calculation. (b) Experimentally obtained
    peak heights as a function of the ac voltage amplitude (measured
    at the microwave source) for $V$ = 13 $\mu$V and $f$ = 52.5 GHz.}
    \label{fig14}
  \end{center}
\end{figure}

Figure \ref{fig14}a shows a calculation of the peak heights as a
function of the ac voltage drop across the barriers $\alpha
=\frac{e \widetilde{V}}{hf}$. Temperature, source-drain voltage,
and frequency are taken from the experiment described below:
$T=200$ mK, $V_{SD}=13$ $\mu $V, and $f=52.5$ GHz. The tunnel
rates from the leads to the ground state and the excited state are
set to $\Gamma_{\varepsilon_0}$ = 5 10$^{8}$ s$^{-1}$ and
$\Gamma_{\varepsilon_1}$ = 14 10$^{8}$ s$^{-1}$, respectively. The
relaxation rate from the excited state to the ground state is
assumed to be zero in the calculation. The effect of a finite
relaxation rate is to reduce the height of $\varepsilon_{1}$ with
respect to the other peaks. The calculated peak heights roughly
follow the Bessel functions in Eq$.$ (\ref{rates}). The
ground-state resonance $\varepsilon_{0}$ follows $J_{0}^{2}(\alpha
)$, since it involves only elastic tunnel events (see Fig$.$
\ref{fig4}, diagram $\varepsilon _{0}$). The photon sidebands
follow $J_{1}^{2}(\alpha )$, since they solely depend on the
probability of photon absorption. For example, the process
$\varepsilon _{0}-hf$ is due to a photon assisted tunnel event
which fills the dot. Once the dot is filled, however, it does not
matter whether the dot is emptied via an elastic or an inelastic
event. The process $\varepsilon _{1}$ follows the product of the
Bessel functions $J_{0}^{2}(\alpha )J_{1}^{2}(\alpha )$ since it
requires that the ground state is emptied via a PAT process, but
also that the following tunneling processes through the excited
state $ \varepsilon _{1}$ are elastic.

Figure \ref{fig14}b shows the experimental results for the peak
heights at $B$ = 0.91 T and $f$ = 52.5 GHz as a function of the ac
voltage amplitude at the output of the source. The measurements
are in good qualitative agreement below an ac source voltage of
100 mV. At higher ac voltages the pumped current starts to become
important. The values for the tunnel rates to $ \varepsilon _{0}$
and to $\varepsilon _{1}$ derived from the dc current-voltage
characteristic are $\Gamma _{\varepsilon_0}$ = 5 10$^{8}$ s$^{-1}$
and $\Gamma_{\varepsilon_1}$ = 6 10$^{8}$ s$^{-1}$. The value for
$\Gamma_{\varepsilon _{1}}$ in the calculation is larger than the
experimentally determined value, but still the calculated value
for the height of the $\varepsilon _1$ resonance is smaller than
the experimental value. It is a general trend in most of our data
that the peak $ \varepsilon _{1}$ is higher than predicted by our
model and that $ \varepsilon _{0}+hf$ is lower than expected from
simulations. A possible explanation has been put forward in Ref$.$
\cite{Brune}.

\subsection{Discussion and Conclusions}
\label{conclusions}

The simulations described earlier, show that the pumped current is
quite independent of the bias voltage when $eV_{SD}\ll hf$, while
current due to the photon resonances increases linearly with the
bias voltage when $ eV_{SD}<k_{B}T$. Therefore, it is possible to
improve the quality of our data by separating the pumped current
from the photon resonances. This could be done by repeating a
measurement at a particular microwave power for different bias
voltages. In addition, the effective electron temperature of 200
mK reported here has now been improved to 50 mK. These two
improvements would allow for a better comparison of future
experiments with calculations over a wider range of microwave
powers.

The conclusion of this work is that photon assisted tunneling is
clearly observed in single-electron transport through small
quantum dots. In addition, microwave irradiation can be used to
perform spectroscopy on the discrete level spectrum. The parameter
dependence of PAT is in reasonable agreement with calculations
based on a master equation. Both, linear-frequency dependence and
Bessel function power dependence are clearly observed. Recently, a
non-equilibrium Green's function method applied to our PAT
studies, has provided very good agreement with our results,
including an explanation for the absence of the sideband at
$\varepsilon _{0}+hf$ \cite{Sun}. Brune {\it et al}. \cite{Brune}
have analyzed the influence of intra-dot transitions. Including
intra-dot transitions, they find good agreement with the height of
the measured $\varepsilon_{1}$ peak in Figs$.$ \ref{fig10} and
\ref{fig11}.

Qualitatively similar results have been obtained on superlattice
structures irradiated by a free-electron laser. The dot and
superlattice experiments have stimulated new theoretical interest
in ac transport through non-superconducting structures. For
instance, new results have been obtained for PAT across a single
tunnel barrier for different sorts of oscillating potentials
\cite{Wagner}; for PAT through superlattices in a self-consistent
treatment \cite{Aguado}, for the ac effects on transport through a
QPC and double barrier structures in a numerical study
\cite{Yakubo}, for the ac effects on correlated transport through
Luttinger liquids \cite{Cuniberti}, and for the Kondo effect in
irradiated quantum dots \cite{Lopez}.

These models and theories are very useful in explaining and
predicting new transport mechanisms. We feel, however, that a word
of caution is appropriate here. Although our PAT experiments give
clean results, we have very little control over the oscillating
potentials inside the sample. We simply couple in a microwave
signal to one of the gates and measure the effect in the dc
current. The dc response is very sensitive to the applied
frequency. Sometimes the microwaves couple in more across the left
barrier and sometimes more across the right barrier. The
sensitivity of the asymmetry in coupling might be due to a
complicated electric field pattern around the metallic gate
structure. We think that to some degree our gate structure acts as
a co-planar waveguide where the two QPC barrier gates serve as
ground planes for the central microwave gate. Even under such
conditions, it is not clear how this oscillating gate potential is
carried over to the electron gas 100 nm below the surface.
Pedersen and B\"{u}ttiker \cite{Pedersen} have recently analyzed
this problem for the case of quantum dots. They stress that
oscillating potentials not only wiggle energy levels but also
generate alternating currents in the sample. The alternating
currents affect the self-consistent potential of the tunnel
barriers and the potential on the dot. In effect, all the voltages
and capacitances become renormalized and depend on the occupation
of the dot. Sideband positions and heights are not expected to
precisely follow a linear frequency dependence nor a Bessel
function behavior. An experimental, quantitative study of the
parameter dependence of the sidebands might give valuable
information about these screening effects. Even more interesting
would be to study the frequency dependence of the current,
including measurements of higher harmonics at multiples of $f$. An
analysis for such a set-up in terms of a frequency dependent
environment is given in Ref$.$ \cite{Aguado2}.

PAT is intrinsically a coherent phenomenon. The PAT measurements
described above, however, are insensitive to the phase of the
transmitted electrons. Coherence in the presence of a
time-dependent field is therefore not directly demonstrated. Jauho
and Wingreen \cite{Jauho} have proposed a PAT measurement through
a quantum dot situated in one of two branches of an Aharonov-Bohm
ring. They find that coherent absorption and reemission of photons
can be detected via a phase measurement at the sidebands. The
proposed mesoscopic double-slit geometry has been successfully
used before to demonstrate coherent transmission through a quantum
dot for the time-{\it independent} case \cite{Yacoby,Schuster}.

By connecting two quantum dots in series, a double quantum dot
system can be obtained. An overview of charging effects in double
quantum dots is given in Ref$.$ \cite{Coulomb}. Theoretical
studies on PAT in a double quantum dot are done by Stoof and
Nazarov \cite{Stoof}, Stafford and Wingreen \cite{Stafford} and
Brune {\it et al}. \cite{Brune1}. PAT studies of a double quantum
dot enable the characterization of the coupling between two
discrete energy levels. Depending on the strength of the inter-dot
coupling, the two dots can form `ionic' or `covalent' bonds. By
varying the inter-dot coupling, Oosterkamp {\it et al. }
\cite{Oosterkamp} experimentally demonstrated the transition from
ionic bonding to covalent bonding in a quantum dot 'artificial
molecule' that is probed by microwave excitations. In the same
frequency regime experiments have been performed by Blick {\it et
al.} \cite {Blick96} and by Fujisawa and Tarucha \cite{Toshi1}.

The study of photon assisted tunneling in quantum dots described
here, forms a valuable extension of the understanding of the dc
properties. Microwave measurements have been proved to be a useful
spectroscopy tool for quantum dot systems. Microwaves are also
expected to play a role of importance in future experiments on
quantum dots and the possible application of quantum dots as solid
state quantum bits. An example would be a time-resolved
measurement of Rabi oscillations in a double quantum dot. In this
case, the microwave signal is used to tune the Rabi oscillation
frequency \cite {Stoof,Stafford}.

We thank R. Aguado, S.F. Godijn, A.E.A. Koolen, J.E.\ Mooij, Yu.V.
Nazarov, R.M. Schouten, T. Fujisawa, S. Tarucha, T.H. Stoof, P.
McEuen and N.C. van der Vaart for experimental help and useful
discussions. This work was supported by the Dutch Organization for
Research on Matter (FOM), by the EU via the TMR network (ERBFMRX
CT98-0180), and by the NEDO joint research program (NTDP-98).

\end{article}

\begin{thebibliography}{99}
\bibitem{B not zero} PAT is also observed at $B=0$ but with less resolution
due to a higher effective electron temperature at $B=0$.
\bibitem{2hf>Deltae} Because $2hf>\Delta \varepsilon $ the excited state
$\varepsilon _{1}$ is weakly visible at $V_{g}\sim -508$ mV in
Fig$.$ 9 where it overlaps with $\varepsilon _{0}+2hf$.
\end{thebibliography}
\end{document}